# 5G New Radio: Unveiling the Essentials of the Next Generation Wireless Access Technology

Xingqin Lin, Jingya Li, Robert Baldemair, Thomas Cheng, Stefan Parkvall, Daniel Larsson, Havish Koorapaty, Mattias Frenne, Sorour Falahati, Asbjörn Grövlen, Karl Werner

Ericsson

Contact: xingqin.lin@ericsson.com;  jingya.li@ericsson.com

*Abstract*— **The 5th generation (5G) wireless access technology, known as new radio (NR), will address a variety of usage scenarios from enhanced mobile broadband to ultra-reliable low-latency communications to massive machine type communications. Key technology features include ultra-lean transmission, support for low latency, advanced antenna technologies, and spectrum flexibility including operation in high frequency bands and inter-working between high and low frequency bands. This article provides an overview of the essentials of the state of the art in 5G wireless technology represented by the 3GPP NR technical specifications, with a focus on the physical layer. We describe the fundamental concepts of 5G NR, explain in detail the design of physical channels and reference signals, and share the various design rationales influencing standardization.**

## I. THE BIRTH OF 3GPP 5G NEW RADIO

The 5th generation (5G) wireless access technology, known as new radio (NR), will address a variety of usage scenarios from enhanced mobile broadband (eMBB) to ultra-reliable low-latency communications (URLLC) to massive machine type communications (mMTC). NR can meet the performance requirements set by the international telecommunication union (ITU) for international mobile telecommunications for the year 2020 (IMT-2020) [1].

The third-generation partnership project (3GPP) is a global standard-development organization and has been developing 5G NR over the past few years. After initial studies [2]-[4], 3GPP approved a work item in March 2017 for NR specifications as part of Release 15 [5]. At the same meeting, 3GPP agreed to a proposal to accelerate the 5G schedule to complete non-standalone (NSA) NR by December 2017, while standalone (SA) NR was scheduled to be completed by June 2018. In NSA operation, long-term evolution (LTE) is used for initial access and mobility handling while the SA version can be deployed independently from LTE. A major milestone was reached in December 2017 with the approval of the NSA NR specifications and the SA version was completed in June 2018. The last step for Rel-15 is a late drop that will be completed by December 2018. The late drop will include more architecture options, e.g., the possibility to connect 5G NodeBs (gNB) to the evolved packet core (EPC) and operating NR and LTE in multi-connectivity mode wherein NR is the master node and LTE is the secondary node.

Key NR features include ultra-lean transmission [6], support for low latency, advanced antenna technologies, and spectrum flexibility including operation in high frequency bands, inter-working between high and low frequency bands, and dynamic time division multiplexing (TDD). A high-level overview of these technology components and NR design principles is provided in [7]. In contrast, the objective of this article is to delve into the detailed NR technical specifications (TS) to unveil the essentials of NR design, while keeping the overall contents at a level accessible for an audience working in the wireless communications and networking communities.

The radio interface of NR consists of the physical layer (Layer 1) and higher layers such as medium access control and radio resource control (RRC). Physical layer specifications are described in the TS 38.200 series [8]-[13], and higher layer specifications are described in the TS 38.300 series (see, e.g. [14] for RRC specifications). We will mainly focus on the physical layer in this article and keep the treatment of higher layers to a minimum.

The remainder of this article is organized as follows. In Section II, we introduce basic NR design and terminologies. In Section III, we describe the synchronization signals (SS), physical broadcast channel (PBCH), and physical random access channel (PRACH). We present the physical downlink shared channel (PDSCH) and physical uplink shared channel (PUSCH) in Section IV, and the physical downlink control channel (PDCCH) and physical uplink control channel (PUCCH) in Section V. The design of reference signals is described in Section VI, followed by our concluding remarks in Section 0.

## II. FUNDAMENTALS OF 5G NEW RADIO

In this section, we present the fundamental concepts of 5G NR design and basic terminologies, with an illustration of the frame structure given in Figure 1.

### A. Waveform, Numerology, and Frame Structure

The choice of radio waveform is the core physical layer decision for any wireless access technology. After assessments of all the waveform proposals, 3GPP agreed to adopt orthogonal frequency division multiplexing (OFDM) with a cyclic-prefix (CP) for both DL and UL transmissions. CP-OFDM can enable low implementation complexity and low cost for wide bandwidth operations and multiple-input multiple-output (MIMO) technologies. NR also supports the use of discrete Fourier transform (DFT) spread OFDM (DFT-S-OFDM) in the uplink to improve coverage.

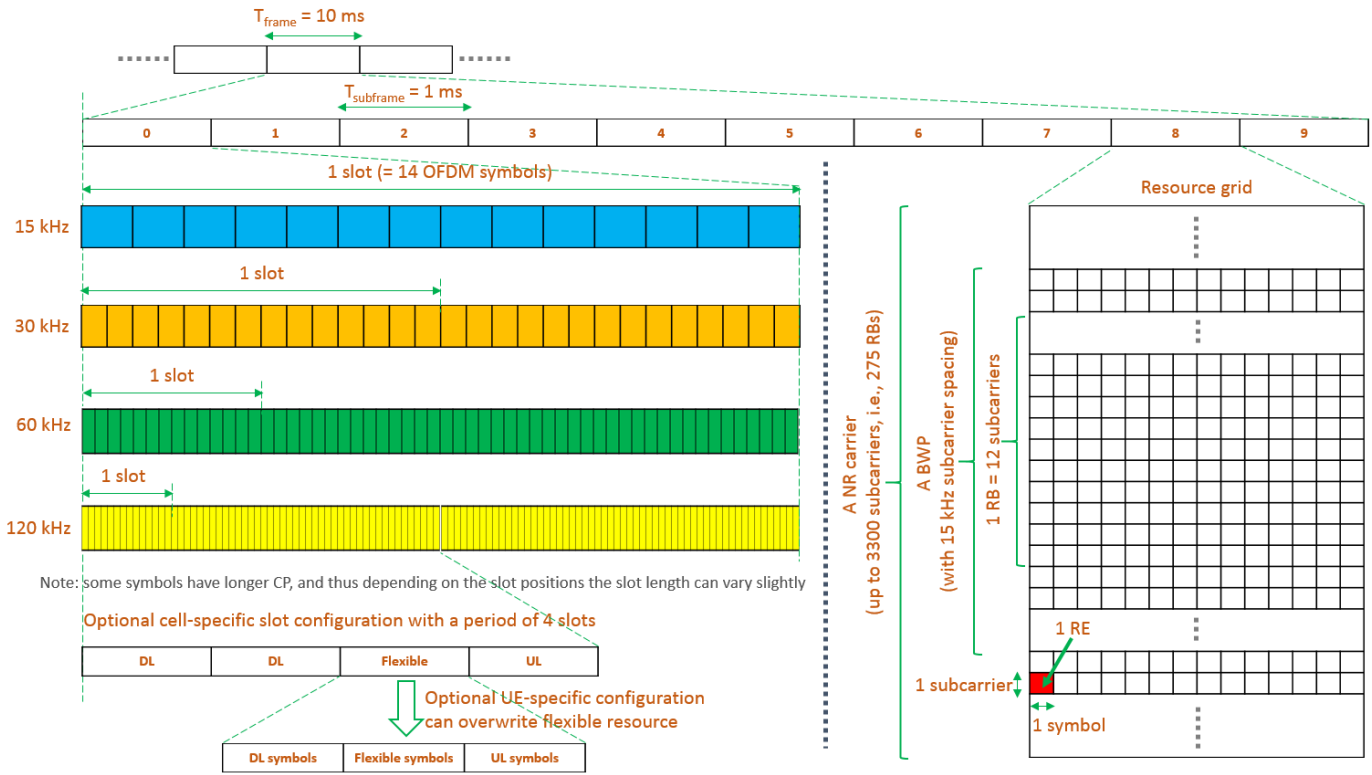

Figure 1: Illustration of 5G NR frame structure and basic terminologies

NR supports operation in the spectrum ranging from sub-1 GHz to millimeter wave bands. Two frequency ranges (FR) are defined in Release 15:
- FR1: 450 MHz – 6 GHz, commonly referred to as sub-6 GHz.
- FR2: 24.25 GHz – 52.6 GHz, commonly referred to as millimeter wave.

Scalable numerologies are key to support NR deployment in such a wide range of spectrum. NR adopts flexible subcarrier spacing of $2^\mu \cdot 15$ kHz ($\mu = 0, 1, \ldots, 4$) scaled from the basic 15 kHz subcarrier spacing in LTE. Accordingly, the CP is scaled down by a factor of $2^{-\mu}$ from the LTE CP length of 4.7 µs. This scalable design allows support for a wide range of deployment scenarios and carrier frequencies. At lower frequencies, below 6 GHz, cells can be larger and subcarrier spacings of 15 kHz and 30 kHz are suitable. At higher carrier frequencies, phase noise becomes more problematic and in FR2, NR supports 60 kHz and 120 kHz for data channels and 120 kHz and 240 kHz for the SS/PBCH block (SSB) used for initial access. At higher frequencies, cells and delay spread are typically smaller and the CP lengths provided by the 60 and 120 kHz numerologies are sufficient.

A frame has a duration of 10 ms and consists of 10 subframes. This is the same as in LTE, facilitating NR and LTE coexistence. Each subframe consists of $2^\mu$ slots of 14 OFDM symbols each. Although a slot is a typical unit for transmission upon which scheduling operates, NR enables transmission to start at any OFDM symbol and last only as many symbols as needed for the communication. This type of "mini-slot" transmission can thus facilitate very low latency for critical data as well as minimize interference to other links per the lean carrier design principle that aims at minimizing transmissions. Latency optimization has been an important consideration in NR. Many other tools besides "mini-slot" transmission have been introduced in NR to reduce latency, as detailed throughout this article.

*B. Resource, Carrier, and Bandwidth Part*

A resource block (RB) consists of 12 consecutive subcarriers in the frequency domain. A single NR carrier in Release-15 is limited to 3300 active subcarriers and to at most 400 MHz bandwidth. The maximum bandwidth in FR1 is 100 MHz, and the maximum bandwidth in FR2 is 400 MHz. Both are much greater than the maximum LTE bandwidth of 20 MHz. Despite wide bandwidth, the ultra-lean design in NR minimizes always-on transmissions, leading to higher network energy efficiency and lower interference.

Operation in millimeter wave bands benefits significantly when complemented by a low frequency carrier to ensure good coverage, especially in the uplink. This can be achieved with the carrier aggregation framework in NR, which is similar to the corresponding framework in LTE. Carrier aggregation provides a tool to combine spectrum in multiple bands. NR supports the possibility to have an NR carrier and an LTE carrier overlapping with each other in frequency, thereby enabling dynamic sharing of spectrum between NR and LTE. This facilitates a smooth migration to NR from LTE. Solutions specified to allow this type of operation are the ability for NR

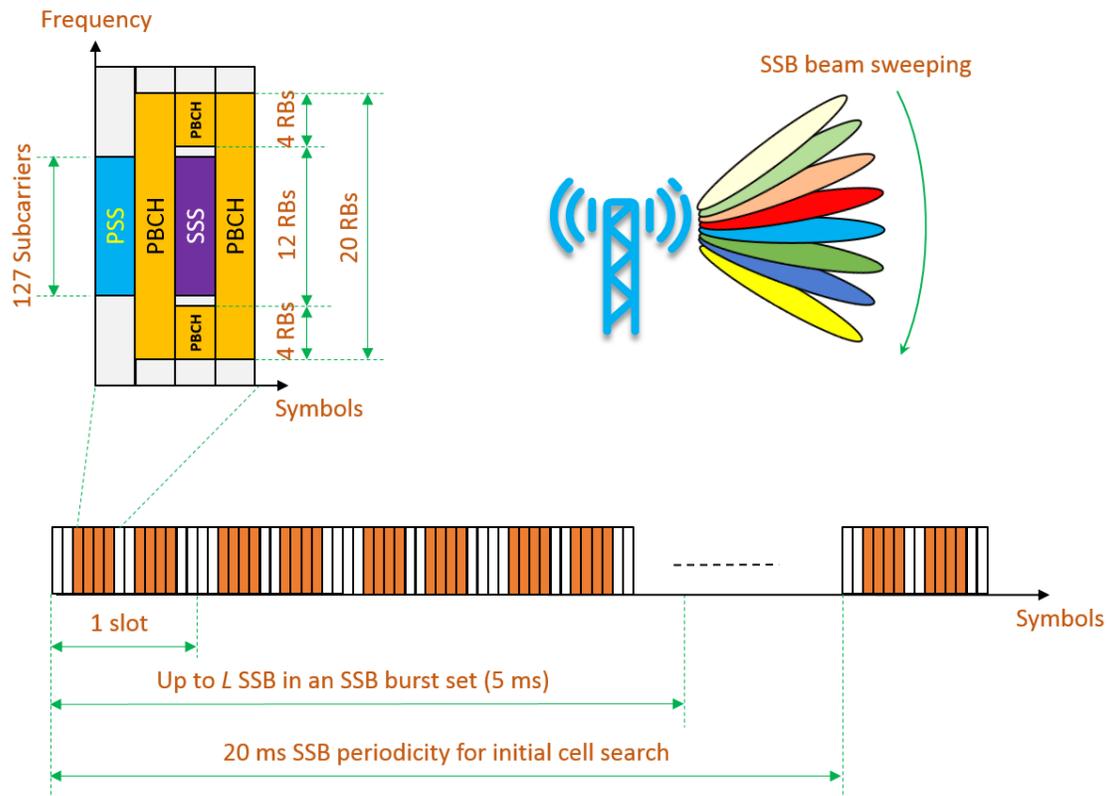

Figure 2: Illustration of 5G NR SS/PBCH

PDSCH to map around LTE cell specific reference signals (CRS), and the possibility of flexible placements of DL control channels, initial access related reference signals and data channels to minimize collisions with LTE reference signals. NR also supports a so-called supplementary uplink (SUL) which can be used as a low-band complement to the cell's UL when operating in high frequency bands and a supplementary DL (SDL), that can be used, for example, in DL only spectrum.

To allow good forward compatibility support in NR, it is possible to configure certain sets of resources to be unused in any PDSCH transmission. This will allow 3GPP to develop new physical layer solutions for currently unknown use cases.

For a carrier with a given subcarrier spacing, the available radio resources in a subframe of duration 1 ms can be thought of as a resource grid composed of subcarriers in frequency and OFDM symbols in time. Accordingly, each resource element (RE) in the resource grid occupies one subcarrier in frequency and one OFDM symbol in time.

To reduce the device power consumption, a user equipment (UE) may be active on a wide bandwidth in case of bursty traffic for a short time, while being active on a narrow bandwidth for the remaining. This is commonly referred to as bandwidth adaptation and is addressed in NR by a new concept known as *bandwidth part*. A bandwidth part is a subset of contiguous RBs on the carrier. Up to four bandwidth parts can be configured in the UE for each of the UL and DL, but at a given time, only one bandwidth part is active per transmission direction. Thus, the UE can receive on a narrow bandwidth part and, when needed, the network can dynamically inform the UE to switch to a wider bandwidth for reception.

*C. Modulation, Channel Coding, and Slot Configuration*

The modulation schemes in NR are similar to LTE, including binary and quadrature phase shift keying (B/QPSK) and quadrature amplitude modulation (QAM) of orders 16, 64, and 256 with binary reflected Gray mapping. NR control channels use Reed-Muller block codes and cyclic redundancy check (CRC) assisted polar codes (versus tail-biting convolutional codes in LTE). NR data channels use rate compatible quasi-cyclic low-density parity-check (LDPC) codes (versus turbo codes in LTE).

The duplexing options supported in NR include frequency division duplex (FDD), TDD with semi-statically configured UL/DL configuration, and dynamic TDD. In TDD spectrum, for small/isolated cells it is possible to use dynamic TDD to adapt to traffic variations, while for large over-the-rooftop cells, semi-static TDD may be more suitable for handling interference issues than fully dynamic TDD.

TDD operations are enabled by flexible slot configuration in NR. Specifically, OFDM symbols in a slot can be configured as 'DL', 'UL', or 'flexible'. DL transmissions can occur in 'DL' or 'flexible' symbols, and UL transmissions can occur in 'UL' or 'flexible' symbols. Cell-specific and UE-specific RRC configurations determine the UL/DL allocations. This framework allows configuration of slot patterns identical to LTE TDD frame structures.

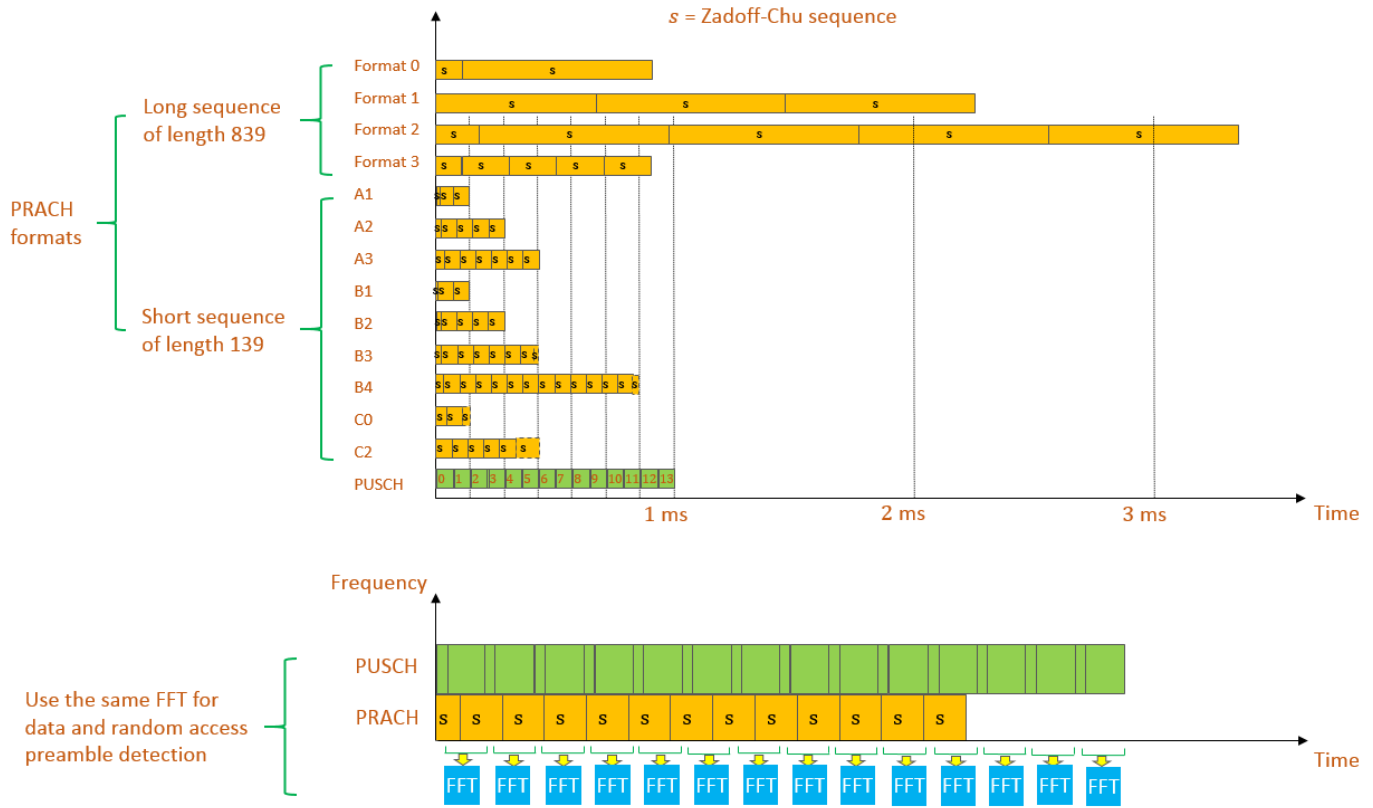

Figure 3: Illustration of 5G NR PRACH

If a slot configuration is not configured, all the resources are considered flexible by default. Whether a symbol is used for DL or UL transmission can be dynamically determined according to layer 1/2 signaling of DL control information (DCI). This leads to a dynamic TDD system.

## III. ACCESS RELATED PHYSICAL CHANNELS AND SIGNALS

### A. Synchronization Signals and Physical Broadcast Channel

Recall that the combination of SS and PBCH is referred to as SSB in NR. The subcarrier spacing of SSB can be 15 or 30 kHz in FR1 and 120 kHz or 240 kHz in FR2. By detecting SS, a UE can obtain the physical cell identity, achieve downlink synchronization in both time and frequency domain, and acquire the timing for PBCH. PBCH carries the very basic system information.

NR SS consists of primary SS (PSS) and secondary SS (SSS). Due to the absence of frequent static reference signals to aid tracking, there could be larger initial frequency errors between the gNB and UEs as compared to LTE, especially for low-cost UEs operating in higher frequencies. To fix the time and frequency offset ambiguity problem of traditional Zadoff-Chu sequence-based LTE PSS, a BPSK modulated m-sequence of length 127 is used for NR PSS. NR SSS is generated by using BPSK modulated Gold sequence of length 127. PSS and SSS together can be used to indicate a total of 1008 different physical cell identities.

An SSB is mapped to 4 OFDM symbols in the time domain and 240 contiguous subcarriers (20 RBs) in the frequency domain, as illustrated in Figure 2. To support beamforming for initial access, a new concept, SS burst set, is introduced in NR to support possible beam sweeping for SSB transmission. To minimize always-on transmissions [6], multiple SSBs are transmitted in a localized burst set in conjunction with a sparse burst set periodicity (default at 20 ms). Within an SS burst set period, up to 64 SSBs can be transmitted in different beams. The transmission of SS blocks within a SS burst set is confined to a 5 ms window. The set of possible SSB time locations within an SS burst set depends on the numerology which in most cases is uniquely identified by the frequency band. The frequency location of SSB is not necessarily in the center of the system bandwidth and is configured by higher layer parameters to support sparser search raster for SSB detection. A sparser raster in frequency is required to compensate for the increased search time due to the sparser SSB periodicity.

### B. Physical Random Access Channel

PRACH is used to transmit a random-access preamble from a UE to indicate to the gNB a random-access attempt and to assist the gNB to adjust the uplink timing of the UE, among other parameters. Like in LTE, Zadoff-Chu sequences are used for generating NR random-access preambles due to their favorable properties, including constant amplitude before and after DFT operation, zero cyclic auto-correlation and low cross-correlation.

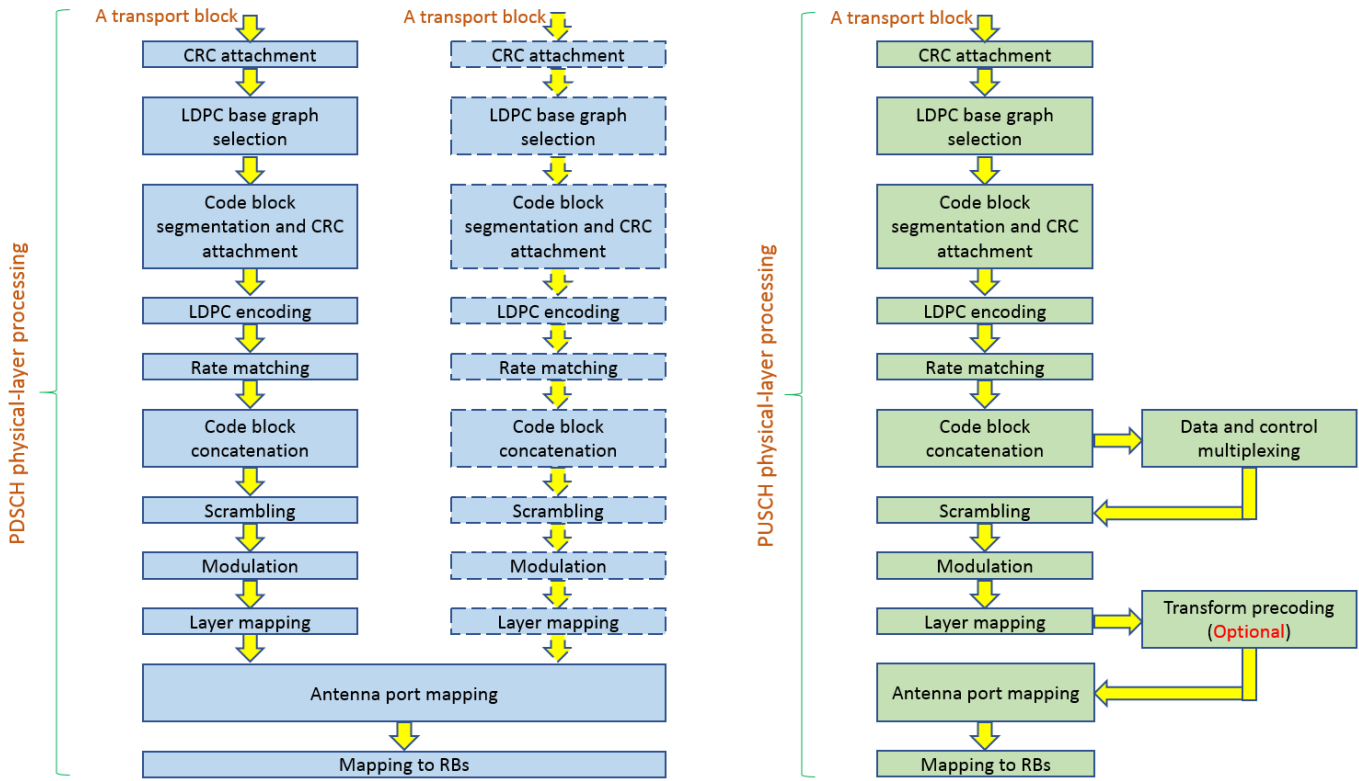

Figure 4: Physical layer processing for 5G NR PDSCH and PUSCH

In contrast to LTE, NR random-access preamble supports two different sequence lengths with different format configurations, as shown in Figure 3, to handle the wide range of deployments for which NR is designed.

For the long sequence of length 839, four preamble formats that originated from the LTE preambles are supported, mainly targeting large cell deployment scenarios. These formats can only be used in FR1 and have a subcarrier spacing of 1.25 or 5 kHz.

For the short sequence of length 139, nine different preamble formats are introduced in NR, mainly targeting the small/normal cell and indoor deployment scenarios. The short preamble formats can be used in both FR1 with subcarrier spacing of 15 or 30 kHz and FR2 with subcarrier spacing of 60 or 120 kHz. In contrast to LTE, for the design of the short preamble formats, the last part of each OFDM symbol acts as a CP for the next OFDM symbol and the length of a preamble OFDM symbol equals the length of data OFDM symbols. There are several benefits of this new design. Firstly, it allows the gNB receiver to use the same fast Fourier transform (FFT) for data and random-access preamble detection. Secondly, due to the composition of multiple shorter OFDM symbols per PRACH preamble, the new short preamble formats are more robust against time varying channels and frequency errors. Thirdly, it supports the possibility of analog beam sweeping during PRACH reception such that the same preamble can be received with different beams at the gNB.

## IV. PHYSICAL SHARED CHANNELS

### A. Physical Downlink Shared Channel

PDSCH is used for the transmission of DL user data, UE-specific higher layer information, system information, and paging.

For transmission of a DL transport block (payload for physical layer), a transport block CRC is first appended to provide error detection, followed by a LDPC base graph selection. NR supports two LDPC base graphs, one optimized for small transport blocks and one for larger transport blocks. Then segmentation of the transport block into code blocks and code block CRC attachment are performed. Each code block is individually LDPC encoded. The LDPC coded blocks are then individually rate matched. Finally, code block concatenation is performed to create a codeword for transmission on the PDSCH. Up to 2 codewords can be transmitted simultaneously on the PDSCH.

The contents of each codeword are scrambled and modulated to generate a block of complex-valued modulation symbols. The symbols are mapped on up to 4 MIMO layers. A PDSCH can have two codewords to support up to 8-layer transmission. The layers are mapped to antenna ports in a specification transparent manner (non-codebook based), hence how beamforming or MIMO precoding operation is performed is up to network implementation and transparent to the UE. For each of the antenna ports (i.e. layers) used for transmission of the PDSCH, the symbols are mapped to RBs.

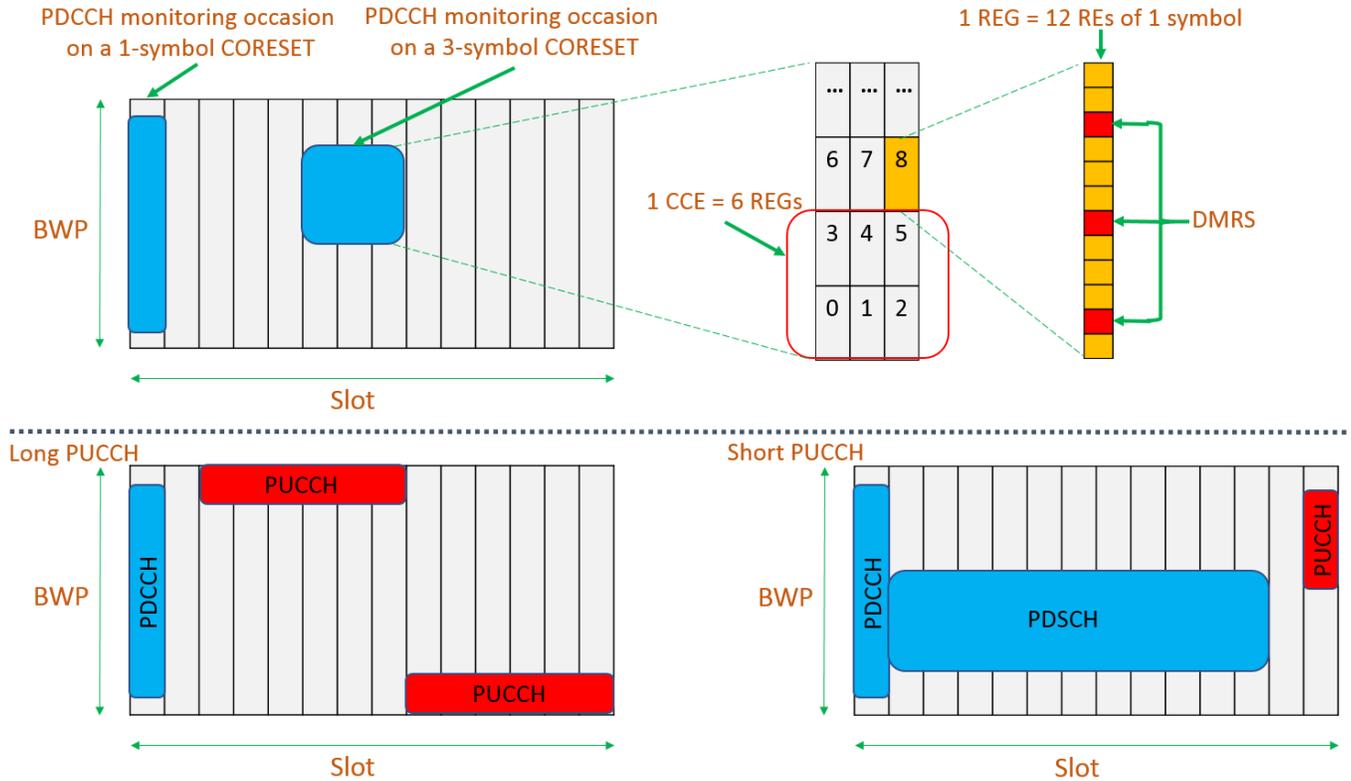

Figure 5: Illustration of 5G NR PDCCH and PUCCH

When receiving unicast PDSCH, UE can be informed that certain resources are not available for PDSCH. These unavailable resources may include configurable rate matching patterns with RB and symbol level granularity or RE level granularity. The latter is used to map around LTE CRS in case NR and LTE share the same carrier. This facilitates both forward and backward capability, since the network can blank radio resources to serve future unknown services while not causing backward compatibility issues.

Physical layer processing for NR PDSCH is summarized in in the left part of Figure 4.

### B. Physical Uplink Shared Channel

PUSCH is used for the transmission of UL shared channel (UL-SCH) and layer 1/2 control information. The UL-SCH is the transport channel used for transmitting an UL transport block. The physical layer processing of an UL transport block is similar to the processing of a DL transport block, as summarized in right part of Figure 4.

The contents of the codeword are scrambled and modulated to generate a block of complex-valued modulation symbols. The symbols are then mapped onto one or several layers. PUSCH supports a single codeword that can be mapped up to 4 layers. In case of a single layer transmission only, DFT transform precoding can optionally be applied if enabled. For the layers to antenna ports mapping, both non-codebook-based transmission and codebook-based transmission are supported in the UL. For each of the antenna ports used for transmission of the physical channel, the symbols are mapped to RBs. In contrast to LTE, the mapping is done in frequency before time to enable early decoding at the receiver.

## V. PHYSICAL CONTROL CHANNELS

### A. Physical Downlink Control Channel

PDCCH is used to carry DCI such as downlink scheduling assignments and uplink scheduling grants. An illustration of NR PDCCH is given in the upper part of Figure 5.

Legacy LTE control channels are always distributed across the entire system bandwidth, making it difficult to control intercell interference [6]. NR PDCCHs are specifically designed to transmit in a configurable control resource set (CORESET). A CORESET is analogous to the control region in LTE but is generalized in the sense that the set of RBs and the set of OFDM symbols in which it is located are configurable with the corresponding PDCCH search spaces. Such configuration flexibilities of control regions including time, frequency, numerologies, and operating points enable NR to address a wide range of use cases.

Frequency allocation in a CORESET configuration can be contiguous or non-contiguous. CORESET configuration in time spans 1-3 consecutive OFDM symbols. The REs in a CORESET are organized in RE groups (REGs). Each REG consists of the 12 REs of one OFDM symbol in one RB. A PDCCH is confined to one CORESET and transmitted with its own demodulation reference signal (DMRS) enabling UE-specific beamforming of the control channel. A PDCCH is carried by 1, 2, 4, 8 or 16 control channel elements (CCEs) to xaccommodate different DCI payload size or different coding rates. Each CCE consists of 6 REGs. The CCE-to-REG mapping for a CORESET can be interleaved (for frequency diversity) or non-interleaved (for localized beam-forming). A UE is configured to blindly monitor a number of PDCCH candidates of different DCI formats and different aggregation levels. The blind decoding processing has an associated UE complexity cost but is required to provide flexible scheduling and handling of different DCI formats with lower overhead.

*B. Physical Uplink Control Channel*

PUCCH is used to carry uplink control information (UCI) such as hybrid automatic repeat request (HARQ) feedback, channel state information (CSI), and scheduling request (SR). An illustration of NR PUCCH is given in the bottom part of Figure 5.

Unlike LTE PUCCH that is located at the edges of the carrier bandwidth and is designed with fixed duration and timing, NR PUCCH is flexible in its time and frequency allocation. That allows supporting UEs with smaller bandwidth capabilities in an NR carrier and efficient usage of available resources with respect to coverage and capacity. NR PUCCH design is based on 5 PUCCH formats. PUCCH formats 0 and 2, a.k.a. short PUCCHs, use 1 or 2 OFDM symbols while PUCCH formats 1, 3 and 4, a.k.a. long PUCCHs, can use 4 to 14 OFDM symbols. PUCCH formats 0 and 1 carry UCI payloads of 1 or 2 bits while other formats are used for carrying UCI payloads of more than 2 bits. In PUCCH formats 1, 3 and 4, symbols with DMRS are time division multiplexed with UCI symbols to maintain low low peak-to-average-power-ratio (PAPR) while in format 2, DMRS is frequency-multiplexed with data-carrying subcarriers. Multi-user multiplexing on the same time and frequency resources is supported only for PUCCH format 0, 1, and 4 by means of different cyclic shifts or OCC when applicable. In the following, additional information on NR PUCCH formats are briefly described:

A UE can be configured with PUCCH resources for CSI reporting or SR. For UCI transmission including HARQ-ACK bits, a UE may be configured with up to 4 PUCCH resource sets based on the UCI size. The first set can only be used for a maximum of 2 HARQ-ACK bits (with a maximum of 32 PUCCH resources) and other sets are applicable for more than 2 bits of UCI (each with a maximum of 8 PUCCH resources). A UE determines the set based on the UCI size, and further indicates a PUCCH resource in the set based on a 3-bit field in DCI (complemented with an implicit rule for the first set with more than 8 resources).

## VI. PHYSICAL REFERENCE SIGNALS

NR reference signal design follows the lean carrier principles [6] – reference signals are on-demand when possible, and their time and frequency distributions are configurable so that requirements can be met with minimal overhead. At low load, reference signal transmission can be extremely sparse. This serves to reduce energy consumption and intercell interference. The high flexibility of the design and the on-demand principle also result in a degree of forward compatibly. In LTE, multiple functions are tied to the always-on CRS. In NR, these functions are supported by multiple UE specifically configured reference signals.

*A. Downlink and Uplink Demodulation Reference Signals (DMRS)*

DMRS is used by the receiver to produce channel estimates for demodulation of the associated physical channel. The design of DMRS is specific for each physical channel – PBCH, PDCCH, PDSCH, PUSCH, and PUCCH. In all cases, DMRS is UE specific, transmitted on demand, and normally does not extend outside of the scheduled physical resource of the channel it supports. In the sequel, we focus on the DMRS for PDSCH and PUSCH when CP-OFDM is used.

The PDSCH/PUSCH DMRS structure supports a wide range of scenarios, UE capabilities, and use cases. The number of DMRS symbols in a PDSCH/PUSCH duration can be configured; this enables support for very high UE mobility, but also low DMRS overhead when the scenario allows so. Similarly, the density of DMRS in the frequency domain is configurable to allow for an optimized overhead. The first DMRS instance comes early in the PDSCH/PUSCH transmission; this enables channel estimation to start early in the receiver, thereby reducing the processing latency. DMRS can be on a regular comb structure, and RB bundling is configurable. This is beneficial for efficient, high performance channel estimation. NR DMRS supports massive multi-user MIMO; it can be beamformed and supports up to 12 orthogonal layers. DMRS sequence for CP-OFDM is QPSK based on Gold sequences. For PUSCH with DFT-S-OFDM there is also a low PAPR Zadoff-Chu mode.

*B. Downlink and Uplink Phase-Tracking Reference Signals (PTRS)*

PTRS is used for tracking the phase of the local oscillator at the receiver and transmitter. This enables suppression of phase noise and common phase error, particularly important at high carrier frequencies such as millimeter wave. Due to the properties of phase noise, PTRS can have low density in the frequency domain but high density in the time domain. PTRS can be present both in the downlink (associated with PDSCH) and in the uplink (associated with PUSCH).

If transmitted, PTRS is always associated with one DMRS port and is confined to the scheduled bandwidth and duration of PDSCH/PUSCH. Time and frequency densities of PTRS are adapted to signal-to-noise-ratio (SNR) and scheduling bandwidth.

*C. Channel-State Information Reference Signals (CSI-RS)*

Similar to the LTE counterpart, NR CSI-RS is used for DL CSI acquisition. Beyond this use case, CSI-RS in NR also supports reference signal received power (RSRP) measurements for mobility and beam management (including analog beamforming), time/frequency tracking for demodulation, and UL reciprocity-based precoding. CSI-RS is UE specifically configured, but multiple user can still share the same resource. Zero-power CSI-RS can be used as a masking tool to protect certain REs by making them unavailable for PDSCH mapping. This masking supports transmission of UE specific CSI-RS, but the design is also a tool for allowing introduction of new features to NR with retained backward compatibility.

NR supports high degree of flexibility for CSI-RS configuration. A resource can be configured with up to 32 ports, and the density is configurable. In the time domain, a CSI-RS resource may start at any OFDM symbol of a slot and it spans 1, 2, or 4 OFDM symbols depending on the number of ports configured. CSI-RS can be periodic, semi-persistent or aperiodic (DCI triggered).

When used for time frequency tracking, CSI-RS can be periodic, or aperiodic. In this use case a single port is configured, and the signal is transmitted in bursts of two or four symbols spread over one or two slots.

*D. Sounding Reference Signals (SRS)*

SRS is used for UL channel sounding. The design supports UL link adaptation and scheduling, but in reciprocity operation also downlink precoder selection, link adaptation and scheduling, e.g., for massive multi-user MIMO.

Contrary to LTE, NR SRS is UE specifically configured. This enables a high degree of flexibility in the system. In the time domain, an SRS resource spans 1, 2 or 4 consecutive symbols mapped within the last 6 symbols of a slot. Multiple SRS symbols allow coverage extension and increased sounding capacity. If multiple resources are configured for a UE, intra-slot antenna switching is also supported (when UE has fewer transmit chains than receive chains). Both these features are important, e.g., in the reciprocity use case. The SRS sequence design and frequency hopping mechanism are similar to LTE SRS.

## VII. CONCLUSIONS

The next generation wireless access technology – 5G NR – serving a wide range of use cases is expected to lead to significant socio-economic benefits. A significant step towards this was achieved when 3GPP approved the highly anticipated standalone 5G NR specifications in June 2018. This article has provided an overview of the essentials of the 3GPP NR specifications representing the state of the art in 5G wireless technology, with a focus on the physical layer. NR is a flexible air interface, capable of meeting a wide range of requirements, use cases, deployments and providing a solid foundation for the future evolution of wireless communications services.